\documentclass{article}
\usepackage{spconf,amsmath,graphicx,hyperref,amssymb}
\usepackage{algorithm,algpseudocode}
\usepackage{xcolor}

\title{Greedy Vector Approximation Based on Rotation for Multiplierless Complex Products}

\name{Luiz F. da S. Coelho$^{\star\dagger}$ \qquad Paulo S. R. Diniz$^\star$ \qquad Didier Le Ruyet$^\dagger$\thanks{This study was financed in part by the Coordena\c{c}\~{a}o de Aperfei\c{c}oamento de Pessoal de N\'{i}vel Superior - Brazil (CAPES) - Finance Code 001, and by CNPq and FAPERJ, both Brazilian research councils. Luiz F. has a CNPq doctoral scholarship. This work was partially supported by the ANR under the France 2030 program, grant NF-PERSEUS ANR-22-PEFT-0004.}}
\address{$^\star$ PEE-COPPE/UFRJ, Rio de Janeiro, Brazil \\ $^\dagger$ CEDRIC-CNAM, Paris, France}

\begin{document}
%
\maketitle
\begin{abstract}
This paper proposes a greedy approximation algorithm for performing complex-valued inner products in multiplierless hardware. The matching pursuits for accurate rotation approximation (MPARA) algorithm approximates the tangent of the phase of complex elements as a sum of signed powers of two (SOPOT) representation, enabling multiplierless hardware for complex-valued multiplication. The proposed algorithm outperforms other state-of-the-art approximation methods for approximating the twiddle factors of a fast Fourier transform (FFT).
\end{abstract}
\section{Introduction}
\label{sec:intro}

Complex-number multiplication is a key operation in many digital signal processing applications for communications. It is the basis for the discrete Fourier transform (DFT), the core of orthogonal frequency division multiplexing (OFDM). The DFT also appears in other countless communication applications, such as multiple-input multiple-output (MIMO) systems, direction of arrival (DoA) estimation, precoding and postcoding in transceivers, and antenna selection~\cite{henriksson2024multi,bae2025finite}.

Performing bit shifts instead of conventional multiplications is a well-known way to reduce the computational cost of inner products. The shift-and-add structure for multiplication falls under multiplierless hardware and is a known design for efficient finite impulse response (FIR) filter hardware designs~\cite{gwee2008low,da2014fir,hartley1996subexpression}.

In~\cite{da2014fir}, the authors explore efficient sums of signed powers of two (SOPOT) representations that rely on signed powers of two (SPTs) to represent real numbers. As the computational complexity is proportional to the number of SPTs, they propose an algorithm called matching pursuits with generalized bit planes (MPGBP) that achieves vector approximation based on SOPOT representations with a minimal number of SPTs for FIR filters. However, their solution is based on real-valued multiplications.

For complex values, multiplication is interpreted as a phase rotation and an amplitude gain. A multiplierless solution for rotation is the coordinate rotation digit computer (CORDIC), which iterates over quantized micro-rotations (MR) performed by bit shifts to approximate a given rotation~\cite{andraka1998survey,meher2009fifty}. Although efficient, CORDIC cannot select the steps it should take in the approximation, so it takes unnecessary steps and leaves room for improvement~\cite{meher2009fifty,garrido2014low,cheng2003ahigh,garrido2025uniformly,jones2026sam,han2026cordic}.

In this paper, we propose a middle-ground algorithm that approximates complex-valued vectors for multiplierless hardware using the fewest bit shifts. The matching pursuits for accurate rotation approximation (MPARA) combines MPGBP and CORDIC. It approximates the vector by considering each element's phase. The solution allocates MRs to the vector elements that most reduce the cost function and enables mixed-precision quantization based on the rotation.

As a use case, we approximate the twiddle factors of a fast Fourier transform (FFT) and show that our method mitigates the non-orthogonality between the transform and its inverse.

\section{Multiplierless Rotation}
\label{sec:}

Let $x,w \in \mathbb{C}$ be two complex values such that $|w| = 1$, where $|\cdot|$ is the absolute value operator. The product of these values is written as
\begin{align}
    y &= xw = x_{\textrm{r}}w_{\textrm{r}} - x_{\textrm{i}}w_{\textrm{i}} + j(x_{\textrm{r}}w_{\textrm{i}} + x_{\textrm{i}}w_{\textrm{r}}) \nonumber \\
    &= y_{\textrm{r}} + j y_{\textrm{i}}, \label{eq:complex-multiplication-cartesian}
\end{align}
where $y \in \mathbb{C}$ is the resulting complex number, $j = \sqrt{-1}$, and $(\cdot)_{\textrm{r}}$ and $(\cdot)_{\textrm{i}}$ indicate the real and imaginary components of the complex number.

\subsection{Complex-Number Multiplication as a Rotation}

\begin{figure}[!t]
    \centering
    \includegraphics[width=0.7\linewidth]{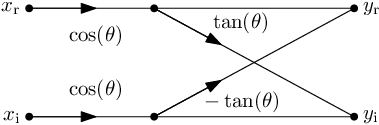}
    \caption{Diagram of complex-number multiplication, $y=xw$, as a phase rotation.}
    \label{fig:diagram}
\end{figure}

Complex numbers have amplitude and phase. When we multiply two complex numbers, the resulting amplitude is the product of the amplitudes, and the resulting phase is the sum of the phases. Let $\theta$ be the phase of $w$, we have
\begin{align}
    y_{\textrm{r}} &= w_{\textrm{r}}\left(x_{\textrm{r}} - x_{\textrm{i}}\frac{w_{\textrm{i}}}{w_{\textrm{r}}}\right) = \cos(\theta) (x_{\textrm{r}} - x_{\textrm{i}}\tan(\theta)) \label{eq:real-part-result} \\
    y_{\textrm{i}} &= w_{\textrm{r}}\left(x_{\textrm{i}} + x_{\textrm{r}}\frac{w_{\textrm{i}}}{w_{\textrm{r}}}\right) = \cos(\theta) (x_{\textrm{i}} + x_{\textrm{r}} \tan(\theta)). \label{eq:imaginary-part-result}
\end{align}
The phase rotation $\theta = \arctan\left(\frac{w_{\textrm{i}}}{w_{\textrm{r}}}\right)$ transforms, by rotation, $x$ into $y$, and by operating \eqref{eq:real-part-result} and \eqref{eq:imaginary-part-result} instead of Eq.~\eqref{eq:complex-multiplication-cartesian}, explains the phase rotation in the cartesian form. Fig.~\ref{fig:diagram} shows the phase rotation, using four multipliers and two adders.

\subsection{Coordinate Rotation Digit Computer (CORDIC)}

CORDIC exploits equations \eqref{eq:real-part-result} and \eqref{eq:imaginary-part-result} by approximating $\tan(\theta)$ as a SOPOT, enabling shift-and-add computation. In each CORDIC iteration, we take an angular step of size $\theta_k = \arctan\left(2^{-k}\right)$ in the direction of the desired rotation, where $k$ is the algorithm's iteration index. Algorithm~\ref{alg:cordic}.

\begin{algorithm}[!t]
    \caption{CORDIC phase rotation}
    \begin{algorithmic}[1]
        \Require{$x\in\mathbb{C}$, $w\in\mathbb{C}$, $k_{\textrm{max}} \in \mathbb{N}$}
        \Ensure{$\hat{y}\in\mathbb{C}$}
        \State $k\gets0$, $\theta_0 \gets \arctan\left(\frac{w_{\textrm{i}}}{w_{\textrm{r}}}\right)$, $y^{(0)} \gets x$ \Comment{Initialization}
        \While{$k < k_{\textrm{max}}$}
            \State $y_{\textrm{r}}^{(k+1)}  \gets y_{\textrm{r}}^{(k)} - \textrm{sign}(\theta_k) y_{\textrm{i}}^{(k)} 2^{-k}$ \Comment{Eq.~\eqref{eq:real-part-result}}
            \State $y_{\textrm{i}}^{(k+1)} \gets y_{\textrm{i}}^{(k)} + \textrm{sign}(\theta_k) y_{\textrm{r}}^{(k)} 2^{-k}$ \Comment{Eq.~\eqref{eq:imaginary-part-result}}
            \State $\theta_{k+1}\gets \theta_k - \textrm{sign}(\theta_k) \arctan(2^{-k})$ \Comment{Eq.~\eqref{eq:phase-update-cordic}}
            \State{$k \gets k + 1$}
        \EndWhile
        \State $\hat{y} \gets y^{(k_{\textrm{max}})}$
     \end{algorithmic}
    \label{alg:cordic}
\end{algorithm}

At each iteration, the residual rotation is calculated as
\begin{align}
    \theta_{k+1} = \theta_k - \textrm{sign}(\theta_k) \arctan\left(2^{-k}\right), \label{eq:phase-update-cordic}
\end{align}
and $\theta_{k+1}$ is the residual rotation and defines the direction of the next iteration's step.

Algorithm~\ref{alg:cordic} disregards the amplitude correction of $\hat{y}$, which is due to $\cos(\theta_k)$ in equations \eqref{eq:real-part-result} and \eqref{eq:imaginary-part-result}. In CORDIC, the correction is performed as follows
\begin{align}
    c = \prod_{k=0}^{k_{\textrm{max}}} \cos(\arctan(2^{-k})) = \prod_{k=0}^{k_{\textrm{max}}} \frac{1}{\sqrt{1+2^{-2k}}},
\end{align}
where $c \in \mathbb{R}$ is the amplitude correction.

As with shift-and-add multiplierless structures, CORDIC hardware design can be optimized to minimize FPGA area, latency, or other criteria~\cite{gwee2008low,hartley1996subexpression,andraka1998survey}.

\subsection{CORDIC and the Signed Bit Plane}


When representing vectors as SOPOT, the vector is mapped into $k_{\textrm{max}}+1$ signed bit planes.
We allocate symbols from the set $\mathcal{M}=\{-1,0,1\}$ in the signed bit plane. A bit plane is defined by its maximum value, or the shallowest bit, and by its minimum value, or deepest bit.

Algorithm~\ref{alg:cordic} allocates symbols from the nonzero set $\mathcal{M}_{\textrm{nz}} = \{-1,1\}$, where $\mathcal{M}_{\textrm{nz}} \subset \mathcal{M}$ in a bit plane. The bit plane starts at $k=0$ and increases in depth in each iteration, up to the maximum bit plane depth, $k_{\textrm{max}}$. Each allocated symbol represents two shift-and-add operations.

By using the set $\mathcal{M}$, we promote sparsity in the bit plane, enabling approximations with few nonzero elements~\cite{da2014fir}. This reduces the number of shift-and-add operations. We propose a greedy algorithm to achieve this approximation.

\section{Greedy Phase Approximation}
\label{sec:rot-mp}

Let ${\bf w} \in \mathbb{C}^N$ be a complex-valued vector, composed as
\begin{align}
    {\bf w} = \begin{bmatrix}
        w(0) & w(1) & \cdots & w(N-1)
    \end{bmatrix}^T,
\end{align}
where $[\cdot]^T$ is the transpose operator, and $w(n) \in \mathbb{C}$. We assume that $|w(n)| = 1$ and that $|\theta_n| < \frac{\pi}{4}$, where $\theta_n$ is the phase of $w(n)$, which can be achieved in a preprocessing step justified further in this section. Let ${\bf r} \in \mathbb{R}^N$ be a vector which elements are $r(n) = \theta_n$, i.e., ${\bf r} = \arctan \left( \Im\{{\bf w}\} \oslash \Re\{{\bf w}\} \right)$, where $\oslash$ is the Hadamard division operator, $\Re\{\cdot\}$ returns the real components of a complex, and $\Im\{\cdot\}$ its imaginary.

Let $\mathcal{D} = \{\boldsymbol{\phi}_p\}_{p\in\Gamma}$ be a dictionary composed of $P$ $N$-length complex-valued vectors, referred as codewords, and $\tilde{\bf r} = \tan({\bf r})$ a vector containing the tangent of each element of ${\bf w}$. The inner product between $\tilde{\bf r}$ and $\boldsymbol{\phi}_p$ is defined as
\begin{align}
    \langle \tilde{\bf r}, \boldsymbol{\phi}_p \rangle = \boldsymbol{\phi}_p^T\tilde{\bf r} = \sum_{n=0}^{N-1} \tilde{r}(n)\phi_p(n),
\end{align}
and the projection of $\tilde{\bf r}$ in $\boldsymbol{\phi}_p$ is
\begin{align}
    \frac{\langle \tilde{\bf r}, \boldsymbol{\phi}_p \rangle}{||\boldsymbol{\phi}_p||_2^2} \boldsymbol{\phi}_p = a_p \boldsymbol{\phi}_p,
\end{align}
where $a_p\in\mathbb{R}$ is the projection coefficient, and $||\cdot||_2$ is the Euclidean norm. We use the following $M$-term approximation
\begin{align}
    {\bf r} \approx \hat{\bf r}_M = \sum_{m=0}^{M-1} \arctan(\underbrace{Q(a_{p_m})}_{2^{-k_m}} ) \boldsymbol{\phi}_{p_m},
\end{align}
where $\hat{\bf r}_M \in \mathbb{R}^N$ is the approximated vector, $Q(\cdot)$ is a quantization function to round $a_{p_m}$ to the nearest power of two, $p_m$ is the $m$-th codeword index, and $k_m$ is the $m$-th bit plane depth.

We base the approximation on a matching pursuits (MP) algorithm, namely the matching pursuits with generalized bit planes (MPGBP)~\cite{da2014fir}. Starting at ${\bf r}_0 = {\bf r}$ and $\hat{\bf r}_0 = {\bf 0}$, the algorithm iteratively performs the approximation by computing
\begin{align}
    {\bf r}_{m+1} &= {\bf r}_m - \arctan\left( 2^{-k_m} \right) \boldsymbol{\phi}_{p_m} \label{eq:residue-update} \\
    \hat{\bf r}_{m+1} &= \hat{\bf r}_m + \arctan\left( 2^{-k_m} \right) \boldsymbol{\phi}_{p_m}, \label{eq:approximation-update}
\end{align}
where $m$ indexes the algorithm iterations. Next, we discuss codeword and bit plane depth selection.

\subsection{Codeword Selection}
The dictionary $\mathcal{D}$ is composed of codewords of length $N$ with elements from the set $\mathcal{M}$, i.e., $\boldsymbol{\phi}_p \in \mathcal{M}^N$. We limit the number of nonzero elements in the codeword to $L \in \mathbb{N}$, with $L < N$. At the $m$-th iteration, we select the codeword that maximizes the projection coefficient,
\begin{align}
    p_m = \arg\max_{p \in \Gamma} \frac{\langle \tilde{\bf r}_m , \boldsymbol{\phi}_p \rangle}{L}, \label{eq:codeword-optimization}
\end{align}
with $L = ||\boldsymbol{\phi}_p||_2^2$. The dictionary size, $|\mathcal{D}| = 2^L \begin{pmatrix} N \\ L \end{pmatrix}$, becomes rapidly impractical for an exhaustive codeword search as $N$ and $L$ increases. However, we can generate the codeword that solves \eqref{eq:codeword-optimization}. Let $\mathcal{N}$ be the set of indices of the vector ${\bf r}_m$. Since the codeword is composed of elements from $\mathcal{M}$, the optimal codeword must select and sum the $L$ greatest absolute values of ${\bf r}_m$, i.e., the set
\begin{align}
    \mathcal{L}^\ast = \arg\max_{\mathcal{L} \subset \mathcal{N}, |\mathcal{L}|=L} \sum_{n \in \mathcal{L}} |\tilde{r}_m(n)|, \label{eq:set-selection}
\end{align}
where $\mathcal{L}$ is any subset of $\mathcal{N}$ with size $L$. Therefore, the codeword is composed as
\begin{align}
    \phi_{p_m}(n) = \begin{cases}
        \textrm{sign}(\tilde{r}_m(n)), &\quad n\in\mathcal{L}^\ast \\
        0, &\quad \textrm{otherwise}
    \end{cases}. \label{eq:codeword-generation}
\end{align}

\subsection{Bit Plane Depth Selection}
At the $m$-th iteration, $Q(\cdot)$ performs the quantization of the projection coefficient, $a_{p_m}$. It consists of finding $k_m$ that best approximates $a_{p_m} \approx 2^{-k_m}$. By bounding $a_{p_m}$ by $2^{-k_m-1}$ and $2^{-k_m + 1}$, we have
\begin{align}
    \frac{2^{-k_m}+2^{-k_m-1}}{2} &\leq \frac{\boldsymbol{\phi}_{p_m}^T\tilde{\bf r}_m}{L} < \frac{2^{-k_m} + 2^{-k_m+1}}{2} \nonumber \\
    k_m &\geq -\log_2 \left( \frac{4\boldsymbol{\phi}_{p_m}^T \tilde{\bf r}_m}{3L}\right) > k_m - 1,
\end{align}
yielding,
\begin{align}
    k_m = \left\lceil -\log_2 \left( \frac{4\boldsymbol{\phi}_{p_m}^T\tilde{\bf r}_m}{3L} \right) \right\rceil. \label{eq:km-calculation}
\end{align}

\subsection{Algorithm and Implementation Constraints}

We propose Algorithm~\ref{alg:mpara} as an MP for accurate rotation approximation (MPARA), summarized in the pseudocode.

\begin{algorithm}[!t]
    \caption{MP for accurate rotation approximation}
    \begin{algorithmic}[1]
        \Require{${\bf w}\in\mathbb{C}^N$, $L\in\mathbb{N}$, $L_{\textrm{max}} \in \mathbb{N}$, $k_{\textrm{max}} \in \mathbb{N}$}
        \Ensure{$\hat{\bf r}_M \in\mathbb{R}^N$, $k_m, \boldsymbol{\phi}_m \forall m$}
        \State $m\gets0$, $\hat{\bf r}_0 \gets {\bf 0}$ \Comment{Initialization} \\ ${\bf r}_0 \gets \arctan( \Im\{{\bf x}\} \oslash \Re\{ {\bf x}\})$ \Comment{Initialization} \\
        $M \gets \left\lfloor \frac{L_{\textrm{max}}}{L} \right\rfloor$ \Comment{Initialization}
        \While{$m < M$}
            \State $\tilde{\bf r}_m \gets \tan({\bf r}_m)$
            \State $\mathcal{L}^\ast \gets \arg \max_{\mathcal{L}\subseteq\mathcal{N},|\mathcal{L}|=L} \sum_{n\in \mathcal{L}} |\tilde{r}_m(n)|$ \Comment{Eq.~\eqref{eq:set-selection}}
            \State $\phi_{m}(n) \gets \begin{cases}
                \textrm{sign}(\tilde{r}_m(n)), &\quad n\in\mathcal{L}^\ast \\
                0, &\quad \textrm{otherwise}
                \end{cases}$ \Comment{Eq.~\eqref{eq:codeword-generation}}
            \State $k_m\gets \left\lceil - \log_{2} \left( \frac{4}{3L} \boldsymbol{\phi}_m^T\tilde{\bf r}_m \right) \right\rceil$ \Comment{Eq.~\eqref{eq:km-calculation}}
            \State $\hat{\bf r}_{m+1} \gets \hat{\bf r}_m + \arctan\left(2^{-k_m} \right)\boldsymbol{\phi}_m$ \Comment{Eq.~\eqref{eq:approximation-update}}
            \State ${\bf r}_{m+1} \gets {\bf r}_m - \arctan\left( 2^{-k_m} \right)\boldsymbol{\phi}_m$ \Comment{Eq.~\eqref{eq:residue-update}}
            \If{$k_m > k_{\textrm{max}}$ \textbf{or} $\epsilon < \frac{{\bf r}_{m+1}^T{\bf r}_{m+1}}{N}$}
                \State $M \gets m$
            \EndIf
            \State $m \gets m + 1$
        \EndWhile
    \end{algorithmic}
    \label{alg:mpara}
\end{algorithm}

\subsubsection{Parameter Considerations}
The algorithm requires four extra parameters, which are described as follows. (i) $L$ specifies the number of non-zero elements in the codewords; it determines how many elements are changed at each iteration. (ii) $L_{\textrm{max}}$ sets the maximum number of non-zero elements in the bit planes; it is a complexity constraint for the final inner product. For a rotation, each non-zero element is equivalent to 2 shifts and 2 additions. (iii) $k_{\textrm{max}}$ sets the maximum bit plane depth, or the maximum length of the bit shifts. (iv) $\epsilon$ sets a minimum approximation error for the total rotation on the vector.

\subsubsection{Input Normalization}
The input signal, ${\bf w} \in \mathbb{C}^N$ is normalized to have elements of amplitude $|w(n)| = 1$ and phase $|\theta|<\frac{\pi}{4}$. In preprocessing, we can normalize the amplitude and phase. Although amplitude normalization might add extra shift-and-add in the final implementation, phase normalization is achieved through trivial rotations of $\pi$ and $\frac{\pi}{2}$ requiring only signal inversions and changes between real and imaginary values. This is mapped in the solution as rotations that require no extra additions.

\subsection{Amplitude Correction}

In CORDIC, the accumulated amplitude gain is calculated as
\begin{align}
    a_{\textrm{gain}} = \lim_{k_{\textrm{max}} \to \infty} \prod_{k=0}^{k_{\textrm{max}}} \sqrt{1 + 2^{-2k}} \approx 1.6468, \label{eq:aplitude-error}
\end{align}
where $a_{\textrm{gain}}$ is the accumulated amplitude gain. Although our proposition uses only a subset of the product in~\eqref{eq:aplitude-error}, the value remains bounded in $\left(1, a_{\textrm{gain}}\right]$. Because rotations in shallower bit planes incur larger amplitude gains, we set the shallowest bit plane to $k=0$ and propose the heuristic in Table~\ref{tab:amplitude-correction} for amplitude correction.
\begin{table}[!t]
    \centering
    \caption{Heuristic for the amplitude gain correction.}
    \begin{tabular}{c||c|c}
        bit plane & Correction & Amplitude Gain \\ \hline\hline
        $0$ & $1 - 2^{-2} - 2^{-4} + 2^{-6}$ & $0.9944$ \\
        $1$ & $1 - 2^{-3} + 2^{-6}$ & $0.9957$ \\
        $2$ & $1 - 2^{-5}$ & $0.9986$ \\
        $k\geq3$ & $1$ & $1.0078 > a_k > 1$ 
    \end{tabular}
    \label{tab:amplitude-correction}
\end{table}

Using a few supplementary shift-and-add operations when bit planes 0, 1, and 2 are active, significantly mitigates amplitude gain. In our proposition, we use no corrections in deeper bit planes, and the amplitude gain is limited to $1.0078$ in the worst-case scenario, which is obtained by activating the depths $k = \left\{ k \in \mathbb{N}: k \neq 0, 1, 2\right\}$. The proposed heuristic incurs a gain in the interval $[0.9887, 1.0078]$.

\section{Results}
\label{sec:results}

Several FFT hardware designs exist~\cite{deng2023radix}, and each benefits from an optimization criterion; however, they still rely on twiddle-factor multiplications. We approximate the twiddle factors of a 256-point FFT. We use the radix-16 FFT for the 256-point FFT and the radix-4 for the 16-point FFT
\begin{align}
    {\bf F}_{256} &= {\bf P}_{256} \left( {\bf I}_{16} \otimes {\bf F}_{16} \right) \textrm{diag}({\bf t}_{256}) \left( {\bf F}_{16} \otimes {\bf I}_{16} \right) \nonumber \\
    {\bf F}_{16} &= {\bf P}_{16} \left({\bf I}_4 \otimes {\bf F}_4 \right) \textrm{diag}( {\bf t}_{16} ) \left( {\bf F}_4 \otimes {\bf I}_{4} \right),
\end{align}
where ${\bf F}_N \in \mathbb{C}^N$ is a $N$-point DFT matrix defined as $\{ {\bf F}_N \}_{k,l} = e^{-j2\pi\frac{kl}{N}}$, $\otimes$ is the Kronecker product, ${\bf I}_N \in \mathbb{R}^{N\times N}$ is the identity matrix, ${\bf t}_N \in \mathbb{C}^N$ is a vector containing the twiddle factors,
\begin{align}
    t_N(n) = e^{-j2\pi \left\lfloor \frac{n}{k} \right\rfloor \frac{n \pmod{k}}{N}},
\end{align}
with $k$ as the radix and ${\bf P}_N \in \mathbb{R}^{N\times N}$ is a permutation matrix defined
\begin{align}
    \{ {\bf P}_N \}_{l,m} = \begin{cases}
        1, &\quad m = k\left\lfloor \frac{l}{k} \right \rfloor + (l \pmod{k}) \\
        0, &\quad \textrm{otherwise}
    \end{cases}.
\end{align}


We select $L=7$ in a binary search for minimal reconstruction phase error degradation, $L_{\textrm{max}}$ is variable so that the number of MR per element varies, and $k_{\textrm{max}} = 6$ is fixed for a maximum bit shift of $6$. We define
\begin{align}
    \hat{\bf J} = \frac{1}{256} \hat{\bf F}_{256}^H \hat{\bf F}_{256},
\end{align}
where $\hat{\bf F}_N \in \mathbb{C}^{N\times N}$ is the approximated FFT matrix, and $\hat{\bf J}\in \mathbb{C}^{N\times N}$. The matrix $\hat{\bf J}$ should approach ${\bf I}_N$ to maintain the transform's property of orthogonality between frequencies. We define $\hat{\bf J}_{\textrm{des}} \in \mathbb{C}^{N\times N}$ as
\begin{align}
    \{ \hat{\bf J}_{\textrm{des}} \}_{l,m} \triangleq \begin{cases}
        \{ \hat{\bf J}_{\textrm{des}} \}_{l,m}, &\quad l=m \\
        0, &\quad \textrm{otherwise}
    \end{cases}.
\end{align}

\begin{figure}[!t]
    \begin{minipage}[b]{.9\linewidth}
        \centering
        \includegraphics[width=\linewidth]{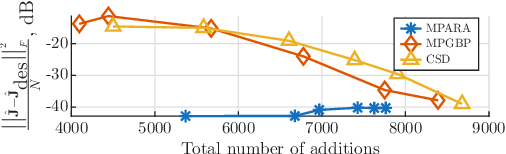}
        \centerline{(a) Average squared error outside the main diagonal.}\medskip
    \end{minipage}
    \begin{minipage}[b]{.9\linewidth}
        \centering
        \includegraphics[width=\linewidth]{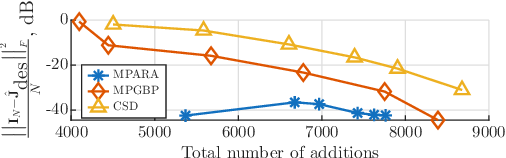}
        \centerline{(b) Average squared error in the main diagonal.}\medskip
    \end{minipage}
    \caption{Study on the orthogonality of the transform.}
    \label{fig:res}
\end{figure}

Figs.~\ref{fig:res}(a) and \ref{fig:res}(b) present the mean squared error outside the main diagonal and on the main diagonal, respectively. These are calculated using the square of the Frobenius norm and averaged by $N$. Our method better approximates the FFT than the other solutions. In our solution, the error increases with the number of additions because of additional MR allocated at bit planes of depth $k\geq3$, which incur additional amplitude gain.

\section{Conclusion}
\label{sec:conclusion}

The paper proposes a greedy complex-valued vector approximation algorithm based on the phase shift, or rotation, of each element, enabling a multiplierless architecture. The MPARA algorithm is a vector approximation method that allocates MRs into elements that most minimize the phase error, yielding approximations with mixed-precision quantization in the phase domain. Our proposition outperforms the CSD and the MPGBP in the approximation of a radix-$16$ $256$-point FFT with radix-$4$ sub-decomposition of the butterflies.



\vfill\pagebreak

\bibliographystyle{IEEEbib}
\bibliography{refs}

\end{document}